\documentclass[10pt, final, journal, letterpaper, twocolumn]{IEEEtran}

\makeatletter
\def\ps@headings{%
\def\@oddhead{\mbox{}\scriptsize\rightmark \hfil \thepage}%
\def\@evenhead{\scriptsize\thepage \hfil \leftmark\mbox{}}%
\def\@oddfoot{}%
\def\@evenfoot{}}
\makeatother 

\IEEEoverridecommandlockouts
\usepackage{cite}

\usepackage{graphicx}
\ifCLASSINFOpdf
\else
\fi
\usepackage{amsmath,amssymb}
\usepackage{algorithm}
\usepackage{algpseudocode}
\usepackage{amsmath}
\usepackage{graphics}
\usepackage{epsfig}

\begin{document}
%
\title{Secure Routing in Multihop Wireless Ad-hoc Networks with Decode-and-Forward Relaying}
%
%
%

\author{\IEEEauthorblockN{Jianping~Yao,
Suili~Feng,
Xiangyun~Zhou,
and
Yuan~Liu
}

\thanks
{
J. Yao, S. Feng, and Y. Liu are with School of Electrical and Information Engineering, South China University of Technology, Guangzhou, P. R. China (e-mails: yaojp\_scut@qq.com, fengsl@scut.edu.cn, eeyliu@scut.edu.cn).

X. Zhou is with the Research School of Engineering, the Australian National University, Canberra, ACT 0200, Australia (e-mail: xiangyun.zhou@anu.edu.au).
}

}

\maketitle

\begin{abstract}
In this paper, we study the problem of secure routing in a multihop wireless ad-hoc network in the presence of randomly distributed eavesdroppers. Specifically, the locations of the eavesdroppers are modeled as a homogeneous Poisson point process (PPP) and the source-destination pair is assisted by intermediate relays using the decode-and-forward (DF) strategy. We analytically characterize the physical layer security performance of any chosen multihop path using the end-to-end secure connection probability (SCP) for both colluding and non-colluding eavesdroppers. To facilitate finding an efficient solution to secure routing, we derive accurate approximations of the SCP. Based on the SCP approximations, we study the secure routing problem which is defined as finding the multihop path having the highest SCP. A revised Bellman-Ford algorithm is adopted to find the optimal path in a distributed manner. Simulation results demonstrate that the proposed secure routing scheme achieves nearly the same performance as exhaustive search.

\end{abstract}

\begin{IEEEkeywords}
Secure connection, physical layer security, multihop routing, homogeneous Poisson point process (PPP), decode-and-forward (DF).
\end{IEEEkeywords}

%
\IEEEpeerreviewmaketitle

\section{Introduction}
%
%
%
%

\IEEEPARstart{N}{etwork} security is a fundamental issue of communication systems. For wireless networks, secure communication is more challenging due to the broadcast nature of wireless channels. The traditional approach for secure communication is to employ the cryptographic algorithms. Recently, physical layer security has emerged as a complementary technology to the cryptography-based method, which can achieve perfect secrecy by properly designing the encoder-decoder of transceivers according to the channel conditions \cite{Wyner1975,Csiszar1978}.

Following the recent advances in cooperative communications, physical layer security in relay networks has captured considerable attention \cite{Dong2010,Wang2012,Mo2014,Mo2012,Zheng2015,Zou2013,Goeckel2011,Saad2012,Sheikholeslami2014,Ding2012,Dehghan2012,Gerbracht2012}. Relay nodes can achieve cooperative diversity by forwarding information or act as cooperative jammers to degrade eavesdroppers' channel conditions, and thus improve the security of legitimate transmission. As an example, the authors in \cite{Dong2010} addressed the secure problem of one source-destination pair with the help of multiple cooperating relays in the presence of one or more eavesdroppers, where three cooperative schemes are considered: decode-and-forward (DF), amplify-and-forward (AF), and cooperative jamming (CJ). The authors in \cite{Wang2012} investigated the distributed beamforming of AF relay network with an external eavesdropper. The authors in \cite{Mo2014} studied the secure beamforming design in a multiple-antenna relay system for maximizing the secrecy sum rate, where the relay is also an internal eavesdropper. The authors in \cite{Mo2012} studied the secure connection probability (SCP) for DF and randomize-and-forward (RaF) relaying strategies where a connection is called secure if the secrecy rate of this connection is positive, as defined in \cite{Zhou2011a}. RaF relaying deviates from the widely-used DF relaying in the way that the relays add independent randomization in each hop when re-encode the received signal \cite{Koyluoglu2012}. The authors in \cite{Zheng2015} performed a comprehensive study on the secure transmission in both DF and RaF two-hop relay networks with only channel distribution information of the wiretap channels, where both the fixed-rate and adaptive-rate transmission at the source and relay nodes were considered. The authors in \cite{Zou2013} derived the intercept probability expressions of optimal relay selection, and the diversity orders of AF and DF were analyzed. The authors in \cite{Goeckel2011} analyzed the relationship between the secrecy performance and the tolerated number of eavesdroppers. The authors in \cite{Saad2012} proposed a tree-formation game to choose secure paths in uplink multihop cellular networks. The authors in \cite{Sheikholeslami2014} considered minimum energy routing in the presence of multiple malicious jammers such that an acceptable end-to-end probability of outage is guaranteed.

A commonly-used assumption in the physical layer security literature is that the channel state information (CSI) or (at least) locations of eavesdroppers are available at legitimate users. To relax such an assumption and take into account the uncertainty of the eavesdroppers' locations, the distribution of the eavesdroppers' locations can be modeled as homogeneous Poisson point processes (PPPs)\cite{Zhou2011,Geraci2014,Tukmanov2014,Cai2014,Wang2013a,Ma2015}. The authors in \cite{Zhou2011} defined the secrecy transmission capacity to study the impact of security requirements on throughput in large-scale decentralized networks consisting of PPP distributed legitimate nodes and eavesdroppers. The authors in \cite{Geraci2014} analyzed the secrecy rates by using regularized channel inversion precoding. The authors in \cite{Tukmanov2014} studied the outage performance based on imperfect CSI. The authors in \cite{Cai2014} proposed a relay selection strategy to improve the SCP, where the locations of both the relays and eavesdroppers follow homogeneous PPPs.

From the above discussions, we find that the problem of secure routing in wireless multihop networks is still largely an open problem. Existing work on secure routing, such as~\cite{Saad2012}, assumed that the CSI or locations of the eavesdroppers are known to the legitimate users, which is often impractical, especially in ad hoc networks enabled by multihop communications. In practice, the eavesdroppers usually work in a passive way, i.e. they just try to overhear as much information as possible conveyed from the legitimate nodes and they do not attempt to actively thwart (i.e. via jamming, signal insertion) the legitimate nodes. In order to statistically characterize the secrecy performance of such scenarios, a PPP is used to statistically model the locations of the eavesdroppers. In this work, we study secure routing in a large-scale multihop wireless network in the presence of randomly-distributed eavesdroppers whose CSI and locations are unknown to the legitimate users. Both colluding and non-colluding eavesdropper scenarios are analyzed. We assume that the intermediate relay nodes use the DF protocol which is the default relaying strategy in wireless ad hoc networks. We assume that the relays use the same codeword as the source, which is a worse-case scenario from security point of view, and hence, is a commonly-used benchmarking scenario in the literature \cite{li2011cooperative,Jeong2011,wang2015hybrid}. With DF relaying, the eavesdroppers can intercept information from multiple hops by maximal-ratio combining. This directly affects the secure routing solutions, e.g., more hops may lead to worse secrecy performance. In passive eavesdropping scenarios, perfect secrecy cannot be guaranteed since the CSI and location of the eavesdroppers are not available at the transmitters. Hence, we adopt the SCP as useful secrecy metrics to characterize the secrecy performance.

The main contributions of this paper are summarized as follows:
\begin{itemize}
  \item For a given path from the source to its destination, we derive exact expressions of SCP for both colluding and non-colluding eavesdroppers, which are used to measure the secrecy performance of that path. Having the exact SCP expressions enables us to analyze and compare the performance of different secure routing solutions.
  \item In order to facilitate finding the secure routing algorithm, we first obtain approximations of the SCP. Based on the SCP approximations, the classical Bellman-Ford routing algorithm is adopted to find the highest SCP path between any given source-to-destination pair in a distributed way.
  \item We conduct simulations to verify the analytical results on SCP and show the effectiveness of the proposed secure routing algorithm. The numerical results show that the proposed secure routing algorithm performs closely to the exhaustive search.
\end{itemize}

The remainder of this paper is organized as follows. In Section \uppercase\expandafter{\romannumeral2}, the system model and performance metric are described. In Section \uppercase\expandafter{\romannumeral3}, the exact expressions of SCP for both colluding and non-colluding eavesdroppers are analyzed. In Section \uppercase\expandafter{\romannumeral4}, we obtain the approximations of SCP, then the secure routing algorithms are derived. In Section \uppercase\expandafter{\romannumeral5}, we present numerical results. Finally, the conclusion is provided in Section \uppercase\expandafter{\romannumeral6}.

\section{System Model and Metric}
We consider a large-scale multihop wireless network with arbitrarily distributed relay nodes and eavesdroppers. We assume that all nodes are equipped with a single omni-directional antenna. An $N$-hop route in the network is a sequence of legitimate nodes $\left(\{{A_i}\},i = 1,\ldots,N + 1\right)$. We assume that every link of the route is exposed to a set of eavesdroppers $\left(\{E_j\},j = 1,2,\ldots\right)$ denoted by ${\Phi _E}$. The eavesdroppers are randomly distributed in the network according to a homogeneous PPP with density ${\lambda _E}$. We assume that the eavesdroppers are passive and thus their CSI as well as locations are unknown to the legitimate nodes. We assume that the legitimate nodes (including source, relay and destination nodes) know the distances between each other. The transmitter of every hop uses a separate slot to transmit the message. We assume that all the channels are modeled by large-scale fading with path loss exponent $\alpha$ along with small-scale Rayleigh fading. Each node $A_{i+1}$ only receives information from its former node $A_{i}$. The instantaneous received signal-to-noise (SNR) at the legitimate node $A_{i+1}$ and eavesdropper $E_{j}$ can be respectively given as
\begin{align}
{{\tt SNR}_{{A_i}{A_{i + 1}}}}\; = \;\frac{{{p_{A_i}}|{h_{{A_i}{A_{i + 1}}}}{|^2}}}{{d_{{A_i}{A_{i + 1}}}^\alpha }},
\end{align}
\begin{align}
{{\tt SNR}_{{A_i}{E_j}}}\; = \;\frac{{{p_{A_i}}|{h_{{A_i}{E_j}}}{|^2}}}{{d_{{A_i}{E_j}}^\alpha }},
\end{align}
 where ${p_{A_i}}$ denotes the transmit power of the legitimate node ${A_i}$; ${d_{{A_i}{A_{i + 1}}}}$ and ${h_{{A_i}{A_{i + 1}}}}$ represent the distance and channel coefficient between nodes ${A_i}$ and ${A_{i + 1}}$, respectively; ${d_{{A_i}{E_j}}}$ and ${h_{{A_i}{E_j}}}$ represent the distance and channel coefficient between nodes ${A_i}$ and ${E_j}$, respectively. We assume that $|{h_{{A_i}{A_{i + 1}}}}{|^2}$ and $|{h_{{A_i}{E_j}}}{|^2}$ follow exponential distributions with mean equal to one. Then according to \cite{Zhou2011a}, the achievable secrecy rate of a single-hop link $A_iA_{i+1}$ is
\begin{align}
{\left[ {\log_2 \left( {1 + {\tt SNR}_{{A_i}{A_{i + 1}}}} \right) - \log_2 \left( {1 +  {{\tt SNR}_{{A_i}{E}}} } \right)} \right]^ + },
\label{PHY_S}
\end{align}
where ${\left[x\right]^ +} = {\max \left({x,0}\right) }$; ${{\tt SNR}_{{A_i}{E}}}$ represents the received SNR at eavesdroppers from the legitimate node ${A_i}$. For the case of non-colluding eavesdroppers, ${{\tt SNR}_{{A_i}{E}}}$ is equivalent to ${\mathop {\max }\limits_{{E_j} \in {\Phi _E}} \left\{ {{\tt SNR}_{{A_i}{E_j}}} \right\}}$, where the maximization operation means the selection of the eavesdropper which has the strongest received signal. For the case of colluding eavesdroppers, ${{\tt SNR}_{{A_i}{E}}}$ is equivalent to $\sum\limits_{{E_j} \in {\Phi _E}} { {{\tt SNR}_{{A_k}{E_j}}} }$.

Due to the multihop DF relaying, we consider that the eavesdroppers can combine the information from multiple hops. Then according to the definition of secure connection in \cite{Zhou2011a}, we say that a given path is secure if the achievable secrecy rate on this path is positive, i.e. (\ref{PHY_S_R}) shown at the top of the next page is satisfied, where ${{I_{{E}}}}$ is the combined received SNR at eavesdroppers from the legitimate transmitter.
\begin{figure*}[!t]
\begin{align}
\frac{1}{N}\left( {\log_2 \left( {1 + \mathop {\min }\limits_{i = 1,\ldots,N} \left\{ {{\tt SNR}_{{A_i}{A_{i + 1}}}} \right\}} \right) - \log_2 \left( {1 + {{I_{{E}}}} } \right)} \right) > 0.
\label{PHY_S_R}
\end{align}
\end{figure*}
 Thus the SCP of a given path can be expressed as (\ref{PHY_S_R_P}), given at the top of the next page.
\begin{figure*}[!t]
\begin{align}
{\mathcal{P}_{DF}} = \mathcal{P}\left( {\frac{1}{N}\left( {\log_2 \left( {1 + \mathop {\min }\limits_{i = 1,\ldots,N} \left\{ {{\tt SNR}_{{A_i}{A_{i + 1}}}} \right\}} \right) - \log_2 \left( {1 + {{I_{{E}}}}} \right)} \right) > 0} \right).
 \label{PHY_S_R_P}
\end{align}
\hrulefill
\end{figure*}

For a given path, ${N}$ is fixed and does not impact the SCP, thus we drop $\frac{1}{N}$ in the analysis on a given path. Then, (\ref{PHY_S_R_P}) can be written as
\begin{align}
{\mathcal{P}_{DF}} = \mathcal{P}\left( {\log_2 \left\{ {\frac{{1 + \mathop {\min }\limits_{i = 1,\ldots,N} \left\{ {{\tt SNR}_{{A_i}{A_{i + 1}}}} \right\}}}{{1 + {{I_{{E}}}}}}} \right\} > 0} \right).
\label{PHY_S_R_P2}
\end{align}

Note that in (\ref{PHY_S_R_P2}), the eavesdropping SNR ${{I_{{E}}}}$ depends on whether the eavesdroppers are colluding or non-colluding. For the case of the non-colluding eavesdroppers, ${{I_{{E}}}}$ is the maximum SNR after MRC among all eavesdroppers (where eavesdropper applies MRC to combine signals received from all hops). For the case of the colluding eavesdroppers, ${{I_{{E}}}}$ is the sum of all SNRs after MRC at all the eavesdroppers. The exact expressions of ${{I_{{E}}}}$ for the both cases will be given in the next section.

\section{Secure Connection Probability of a Given Path}
In this section, we derive the exact SCP of a given path for both colluding and non-colluding eavesdroppers.
\subsection{SCP for Colluding Eavesdroppers}
For the colluding case, the eavesdroppers can share their eavesdropped information. In this case, all the information obtained by the eavesdroppers can be combined, which is the worst scenario from the security point of view. The combined received SNR at eavesdroppers from all hops is given as
\begin{eqnarray}
\nonumber{I_{{E}\_C}} &=& \sum\limits_{{E_j} \in {\Phi _E}} {\sum\limits_{k = 1}^N {{\tt SNR}_{{A_k}{E_j}}} } \\
 &=& \sum\limits_{{E_j} \in {\Phi _E}} {\sum\limits_{k = 1}^N {\frac{{{p_{{A_k}}}{{\left| {{h_{{A_k}{E_j}}}} \right|}^2}}}{{d_{{A_k}{E_j}}^\alpha }}} }.
\label{SUM_SNR_E}
\end{eqnarray}
Then the SCP in (\ref{PHY_S_R_P2}) can be rewritten as
\begin{align}
{\mathcal{P}_{C}}= \mathcal{P}\left( {\log_2 \left( {\frac{{1 + \mathop {\min }\limits_{i = 1,\ldots,N} \left\{ {\frac{{{p_{{A_i}}}{{\left| {{h_{{A_i}{A_{i + 1}}}}} \right|}^2}}}{{d_{{A_i}{A_{i + 1}}}^\alpha }}} \right\}}}{{1 + \sum\limits_{{E_j} \in {\Phi _E}} {\sum\limits_{k = 1}^N {\frac{{{p_{{A_k}}}{{\left| {{h_{{A_k}{E_j}}}} \right|}^2}}}{{d_{{A_k}{E_j}}^\alpha }}} } }}} \right) > 0} \right),
\label{DF_C_1}
\end{align}
which is equivalent to (\ref{DF_C_1_1}), shown at the top of the next page.
\begin{figure*}[!t]
\begin{align}
{\mathcal{P}_{C}}= \mathcal{P}\left( {\mathop {\min }\limits_{i = 1,\ldots,N} \left\{ {\frac{{p_{{A_i}}}{{{\left| {{h_{{A_i}{A_{i + 1}}}}} \right|}^2}}}{{d_{{A_i}{A_{i + 1}}}^\alpha }}} \right\} > \sum\limits_{{E_j} \in {\Phi _E}} {\sum\limits_{k = 1}^N {\frac{{p_{{A_k}}}{{{\left| {{h_{{A_k}{E_j}}}} \right|}^2}}}{{d_{{A_k}{E_j}}^\alpha }}} } } \right).
\label{DF_C_1_1}
\end{align}
\end{figure*}

Since each $|{h_{{A_i}{A_{i + 1}}}}{|^2}$ is independent exponentially distributed random variable with unit mean and independent of ${\Phi _E}$, and $\mathop {\min }\limits_{i = 1,\ldots,N} \left\{ {\frac{{p_{{A_i}}}{{{\left| {{h_{{A_i}{A_{i + 1}}}}} \right|}^2}}}{{d_{{A_i}{A_{i + 1}}}^\alpha }}} \right\}$ is also exponentially distributed with the mean of $\sum\limits_{i = 1}^N {\frac{{d_{{A_i}{A_{i + 1}}}^\alpha }}{p_{{A_i}}}}$. Then (\ref{DF_C_1_1}) can be derived as (\ref{DF_C_2}), written at the top of the next page,
\begin{figure*}[!t]
\begin{eqnarray}
\nonumber{\mathcal{P}_{C}} &=& {{\mathbb{E}}_{\mathop {{{{h}}_{{A_k}{{{E}}_{{j}}}}}}\limits_{k = 1,\ldots,N} ,{\Phi _E}}}\left\{ {{{\exp}}\left[ {{{ - }}\left( {\sum\limits_{i = 1}^N {\frac{{d_{{A_i}{A_{i + 1}}}^\alpha }}{p_{{A_i}}}} } \right)\left( {\sum\limits_{{E_j} \in {\Phi _E}} {\sum\limits_{k = 1}^N {\frac{{p_{{A_k}}}{{{\left| {{h_{{A_k}{E_j}}}} \right|}^2}}}{{d_{{A_k}{E_j}}^\alpha }}} } } \right)} \right]} \right\}\\
\quad \;{\kern 1pt} {\kern 1pt} &\mathop  = \limits^\eta &  {{\mathbb{E}}_{{\Phi _E}}}\left\{ {\prod\limits_{{E_j} \in {\Phi _E}} {\prod\limits_{{{k}} = {{1}}}^N {\frac{{{1}}}{{1 + \frac{p_{{A_k}}}{{d_{{A_k}{E_j}}^{\alpha }}}{\sum\limits_{i = 1}^N {\frac{{d_{{A_i}{A_{i + 1}}}^\alpha }}{p_{{A_i}}}} } }}} } } \right\}.\label{DF_C_2}
\end{eqnarray}
\hrulefill
\end{figure*}
 where the last step $\eta$ stands as ${\left| {{h_{{A_k}{E_j}}}} \right|^2}$ is independent and identically distributed, thus the expectation over the sum of ${\left| {{h_{{A_k}{E_j}}}} \right|^2}$ is equal to the product of the expectation over ${\left| {{h_{{A_k}{E_j}}}} \right|^2}$. For a homogeneous PPP, the probability generating functional (PGFL) is given by \cite{Chiu2013}
\begin{align}
{\mathbb{E}_{\Phi _E}}\left[ {\prod\limits_{{E_j} \in {\Phi _E} } {f\left( x_{E_j} \right)} } \right] =
 \exp \left[ { - \lambda_E \int_{{\mathbb{R}^2}} { {1 - f\left( x_{E_j} \right)} dx_{E_j}} } \right],
\label{PGFL}
\end{align}
where $x_{E_j}$ is the location of ${E_j}$.

Then (\ref{DF_C_2}) can be turned to
\begin{align}
{\mathcal{P}_{C}} = {{\exp}}\left[ {{{ - }}{\lambda _{{E}}}\int_{{\mathbb{R}}^{{2}}} {{{1-}}\prod\limits_{{{k}} = {{1}}}^N {\frac{{{1}}}{{1 + \frac{p_{{A_k}}}{{d_{{A_k}{E_j}}^{\alpha }}}{\sum\limits_{i = 1}^N {\frac{{d_{{A_i}{A_{i + 1}}}^\alpha }}{p_{{A_i}}}} } }} }  dx_{E_j}} } \right].
\label{DF_C_3}
\end{align}

\subsection{SCP for Non-Colluding Eavesdroppers}
In this subsection, we analyze the SCP for non-colluding eavesdroppers. In this case, the eavesdroppers are non-cooperative, so the performance is limited by the eavesdropper which has the strongest received signal. The combined received SNR at eavesdroppers from all hops can be written as
\begin{eqnarray}
\nonumber{I_{{E}\_N}} &=& \mathop {\max }\limits_{{E_j} \in {\Phi _E}} \left\{\sum\limits_{k = 1}^N {{\tt SNR}_{{A_k}{E_j}}}\right\} \\
 &=& \mathop {\max }\limits_{{E_j} \in {\Phi _E}} \left\{\sum\limits_{k = 1}^N {\frac{{{p_{{A_k}}}{{\left| {{h_{{A_k}{E_j}}}} \right|}^2}}}{{d_{{A_k}{E_j}}^\alpha }}}\right\}.
\end{eqnarray}
As the case of colluding eavesdroppers, we also obtain an exact expression of the SCP under the case of non-colluding eavesdroppers. Similar to (\ref{DF_C_1}), we can define the SCP for the case of non-colluding eavesdroppers as (\ref{DF_N_1}) at the next page.
\begin{figure*}[!t]
\begin{eqnarray}
\nonumber{\mathcal{P}_{N}} &=& \mathcal{P}\left( {\log_2 (\frac{{1 + \mathop {\min }\limits_{i = 1,\ldots,N} \left\{ {\frac{{{p_{{A_i}}}{{\left| {{h_{{A_i}{A_{i + 1}}}}} \right|}^2}}}{{d_{{A_i}{A_{i + 1}}}^\alpha }}} \right\}}}{{1 + \mathop {\max }\limits_{{E_j} \in {\Phi _E}} \left\{ {\sum\limits_{k = 1}^N {\frac{{{p_{{A_k}}}{{\left| {{h_{{A_k}{E_j}}}} \right|}^2}}}{{d_{{A_k}{E_j}}^\alpha }}} } \right\}}}) > 0} \right)\\
\quad \;\, &=& \mathcal{P}\left( {\mathop {\min }\limits_{i = 1,\ldots,N} \left\{ {\frac{{{p_{{A_i}}}{{\left| {{h_{{A_i}{A_{i + 1}}}}} \right|}^2}}}{{d_{{A_i}{A_{i + 1}}}^\alpha }}} \right\} > \mathop {\max }\limits_{{E_j} \in {\Phi _E}} \left\{ {\sum\limits_{k = 1}^N {\frac{{{p_{{A_k}}}{{\left| {{h_{{A_k}{E_j}}}} \right|}^2}}}{{d_{{A_k}{E_j}}^\alpha }}} } \right\}} \right).
\label{DF_N_1}
\end{eqnarray}
\end{figure*}

Since ${\Phi _E}$ is a homogeneous PPP, (\ref{DF_N_1}) can be turned to (\ref{DF_N_2}), shown at the next page.
\begin{figure*}[!t]
\begin{align}
{\mathcal{P}_{N}} = {{\mathbb{E}}_{\mathop {{{{h}}_{{A_i}{A_{i + 1}}}}}\limits_{i = 1,\ldots,N} ,{\Phi _E}}}\left\{ {\prod\limits_{{E_j} \in {\Phi _E}} {\mathcal{P}\left( {\mathop {\min }\limits_{i = 1,\ldots,N} \left\{ {\frac{{{p_{{A_i}}}{{\left| {{h_{{A_i}{A_{i + 1}}}}} \right|}^2}}}{{d_{{A_i}{A_{i + 1}}}^\alpha }}} \right\} > \sum\limits_{k = 1}^N {\frac{{{p_{{A_k}}}{{\left| {{h_{{A_k}{E_j}}}} \right|}^2}}}{{d_{{A_k}{E_j}}^\alpha }}} \Bigg| {\mathop {{{{h}}_{{A_i}{A_{i + 1}}}}}\limits_{i = 1,\ldots,N} ,{\Phi _E}} } \right)} } \right\}.
\label{DF_N_2}
\end{align}
\end{figure*}
Then using (\ref{PGFL}) and (\ref{DF_N_2}), we can get (\ref{DF_N_3}), presented at the next page.
\begin{figure*}[!t]
\begin{align}
{\mathcal{P}_{N}} = {{\mathbb{E}}_{\mathop {{{{h}}_{{A_i}{A_{i + 1}}}}}\limits_{i = 1,\ldots,N} }}\left\{ {{{\exp}}\left[ {{{ - }}{\lambda _{{E}}}\int_{{{{\mathbb{R}}}^{{2}}}} {\mathcal{P}\left( {\mathop {\min }\limits_{i = 1,\ldots,N} \left\{ {\frac{{{p_{{A_i}}}{{\left| {{h_{{A_i}{A_{i + 1}}}}} \right|}^2}}}{{d_{{A_i}{A_{i + 1}}}^\alpha }}} \right\}< \sum\limits_{k = 1}^N {\frac{{{p_{{A_k}}}{{\left| {{h_{{A_k}{E_j}}}} \right|}^2}}}{{d_{{A_k}{E_j}}^\alpha }}} \Bigg| {\mathop {{{{h}}_{{A_i}{A_{i + 1}}}}}\limits_{i = 1,\ldots,N} } }\right)}d{x_{{E_j}}} } \right]} \right\}.
\label{DF_N_3}
\end{align}
\end{figure*}

According to \cite{Amari1997}, for a set of independent exponential random variables ${X} = \left\{ {{X_1}, \ldots ,{X_n}} \right\}$ with the parameters of ${\lambda _{{X_i}}},i = 1, \ldots ,n$, the cumulative distribution function (CDF) of the sum of independent not identical exponentially distributed random variables $Y = \sum\limits_{i = 1}^n {{X_i}}$ is given by
\begin{align}
\mathcal{P}\left\{ {Y < y} \right\} = \sum\limits_{i = 1}^n {{\delta _i}\left( {1 - \exp \left[ { - {\lambda _{{X_i}}}y} \right]} \right)},
\label{SUM_EXP}
 \end{align}
where
\begin{align}
{\delta _i} = \prod\limits_{j = 1,j \ne i}^n {\frac{{{\lambda _{{X_j}}}}}{{{\lambda _{{X_j}}} - {\lambda _{{X_i}}}}}}.
\label{DELTA}
\end{align}

Basing on (\ref{DF_N_3}) and (\ref{SUM_EXP}), we obtain the SCP in (\ref{DF_N_4}) shown at the next page.
\begin{figure*}[!t]
\begin{align}
{\mathcal{P}_{N}} = {{\mathbb{E}}_{\mathop {{{{h}}_{{A_i}{A_{i + 1}}}}}\limits_{i = 1,\ldots,N} }}\left\{ {{{\exp}}\left[ {{{ - }}{\lambda _{{E}}}\int_{{{{\mathbb{R}}}^{{2}}}} {\sum\limits_{k = 1}^N {\prod\limits_{m = 1,m \ne k}^N {\frac{{p_{{A_m}}^{ - 1}d_{{A_m}{E_j}}^\alpha }}{{p_{{A_m}}^{ - 1}d_{{A_m}{E_j}}^\alpha  - p_{{A_k}}^{ - 1}d_{{A_k}{E_j}}^\alpha }}} \exp \left[ { - \frac{d_{{A_k}{E_j}}^\alpha}{p_{{A_k}}} \mathop {\min }\limits_{i = 1,\ldots,N} \left\{ {\frac{{{p_{{A_i}}}{{\left| {{h_{{A_i}{A_{i + 1}}}}} \right|}^2}}}{{d_{{A_i}{A_{i + 1}}}^\alpha }}} \right\}} \right]} d{x_{{E_j}}}} } \right]} \right\}.
\label{DF_N_4}
\end{align}
\hrulefill
\end{figure*}

\section{Routing Algorithm}
In Section \uppercase\expandafter{\romannumeral3}, we derived the exact expressions of the SCP under the cases of colluding and non-colluding eavesdroppers for any given path. In this section, we obtain approximations of the SCP to facilitate finding the secure routing algorithm. The approximations are shown to be close to the exact SCP in Section \uppercase\expandafter{\romannumeral5}. The simple analytical form of the SCP approximations allows us to derive efficient secure routing algorithms.

\subsection{Approximation of SCP for Colluding Eavesdroppers}
\newtheorem{lemma}{Lemma}
\begin{lemma}\label{lemma1}
Let ${a_k}\left( {k = 1,2, \ldots,n} \right)$ be arbitrary constants and ${B_k}\left( {k = 1,2, \ldots,n} \right)$ be arbitrary non-negative constants. Let $a$ be anyone of $\{a_k\}$. For an arbitrary positive integer $n$,
\begin{align}
\nonumber\int_{ - \infty }^\infty  {\left( {{{1 - }}\prod\limits_{{{k}} = {{1}}}^n {\frac{{{1}}}{{1 + {B_k}{{(x + {a_k})}^{ - 2}}}}} } \right)} dx\\
 \ge \int_{ - \infty }^\infty  {\left( {{{1 - }}\prod\limits_{{{k}} = {{1}}}^n {\frac{{{1}}}{{1 + {B_k}{{(x + {a})}^{ - 2}}}}} } \right)} dx.
\end{align}
\begin{proof}
See Appendix \ref{appendices_A}.
\end{proof}
\end{lemma}
\vspace{3ex}

Applying Lemma \ref{lemma1}, we can obtain an upper bound of (\ref{DF_C_3}) given in (\ref{DF_C_upper}) where $A$ can be anyone of $\{A_k\}$, shown at the next page. Then we use the upper bound (\ref{DF_C_upper}) as an approximation of (\ref{DF_C_3}).
\begin{figure*}[!t]
\begin{align}
{\mathcal{P}_{C\_approx}} = {{\exp}}\left[ {{{ - }}{\lambda _{{E}}}\int_{{{\mathbb{R}}^{{2}}}} {\left( {{{1 - }}\prod\limits_{{{k}} = {{1}}}^N {\frac{{{1}}}{{1 + \frac{{{p_{{A_k}}}}}{{d_{{A}{E_j}}^\alpha }}\left( {\sum\limits_{i = 1}^N {\frac{{d_{{A_i}{A_{i + 1}}}^\alpha }}{{{p_{{A_i}}}}}} } \right)}}} } \right)d{x_{{E_j}}}} } \right].
\label{DF_C_upper}
\end{align}
\hrulefill
\end{figure*}

\subsection{Approximation of SCP for Non-Colluding Eavesdroppers}
In Eq. (\ref{DF_N_4}), we derived the exact expression of SCP for the case of non-colluding eavesdroppers, which is applicable for all conditions of legitimate nodes' and eavesdroppers' densities. However, the exact expression has a complex form and involves a mathematical expectation. To facilitate finding an efficient solution to the secure routing problem, we resort to an approximation of the SCP. We derive the approximation by considering that the set of legitimate nodes are assumed to share identical distance from an arbitrary eavesdropper. Based on the assumption, we can obtain an approximation of (\ref{DF_N_3}) as (\ref{DF_N_approx1}), given at the top of the page.
\begin{figure*}[!t]
\begin{eqnarray}
\nonumber{\mathcal{P}_{N\_approx1}} &=& {{\mathbb{E}}_{\mathop {{{{h}}_{{A_i}{A_{i + 1}}}}}\limits_{i = 1,\ldots,N} }}\left\{ {{{\exp}}\left[ {{{ - }}{\lambda _{{E}}}\int_{{{{\mathbb{R}}}^{{2}}}} {\mathcal{P}\left( {\mathop {\min }\limits_{i = 1,\ldots,N} \left\{ {\frac{{{p_{{A_i}}}{{\left| {{h_{{A_i}{A_{i + 1}}}}} \right|}^2}}}{{d_{{A_i}{A_{i + 1}}}^\alpha }}} \right\}< \sum\limits_{k = 1}^N {p_{{A_k}}}{\frac{{{{\left| {{h_{{A}{E_j}}}} \right|}^2}}}{{d_{{A}{E_j}}^\alpha }}} \Bigg| {\mathop {{{{h}}_{{A_i}{A_{i + 1}}}}}\limits_{i = 1,\ldots,N} } }\right)d{x_{{E_j}}} } } \right]} \right\}\\
&=& {{\mathbb{E}}_{\mathop {{{{h}}_{{A_i}{A_{i + 1}}}}}\limits_{i = 1,\ldots,N} }}\left\{ {{{\exp}}\left[ {{{ - }}{\lambda _{{E}}}\int_{{{{\mathbb{R}}}^{{2}}}} { { \exp \left[ { - \frac{d_{{A}{E_j}}^\alpha}{\sum\limits_{k = 1}^N {p_{{A_k}}}} \mathop {\min }\limits_{i = 1,\ldots,N} \left\{ {\frac{{{p_{{A_i}}}{{\left| {{h_{{A_i}{A_{i + 1}}}}} \right|}^2}}}{{d_{{A_i}{A_{i + 1}}}^\alpha }}} \right\}} \right]} d{x_{{E_j}}}} } \right]} \right\}.
\label{DF_N_approx1}
\end{eqnarray}
\hrulefill
\end{figure*}

Using Jensen¡¯s inequality, (\ref{DF_N_approx1}) can be turned to (\ref{DF_N_approx2}), given at the next page.
\begin{figure*}[!t]
\begin{align}
{\mathcal{P}_{N\_approx2}} = {{{\exp}}\left[ {{\mathbb{E}}_{\mathop {{{{h}}_{{A_i}{A_{i + 1}}}}}\limits_{i = 1,\ldots,N} }}\left\{{{{ - }}{\lambda _{{E}}}\int_{{{{\mathbb{R}}}^{{2}}}} { { \exp \left[ { - \frac{d_{{A}{E_j}}^\alpha}{\sum\limits_{k = 1}^N {p_{{A_k}}}} \mathop {\min }\limits_{i = 1,\ldots,N} \left\{ {\frac{{{p_{{A_i}}}{{\left| {{h_{{A_i}{A_{i + 1}}}}} \right|}^2}}}{{d_{{A_i}{A_{i + 1}}}^\alpha }}} \right\}} \right]} d{x_{{E_j}}}} } \right\}\right]}.
\label{DF_N_approx2}
\end{align}
\end{figure*}

Since $\mathop {\min }\limits_{i = 1,\ldots,N} \left\{ {\frac{{p_{{A_i}}}{{{\left| {{h_{{A_i}{A_{i + 1}}}}} \right|}^2}}}{{d_{{A_i}{A_{i + 1}}}^\alpha }}} \right\}$ is exponentially distributed with the mean of $\sum\limits_{i = 1}^N {\frac{{d_{{A_i}{A_{i + 1}}}^\alpha }}{p_{{A_i}}}}$, then (\ref{DF_N_approx2}) can be derived as (\ref{DF_N_approx3}), written at the top of the next page.
\begin{figure*}[!t]
\begin{align}
{\mathcal{P}_{N\_approx2}} = \exp \left[ { - {\lambda _E}\int_{{\mathbb{R}^2}} {\frac{{\sum\limits_{i = 1}^N {p_{{A_i}}^{ - 1}d_{{A_i}{A_{i + 1}}}^\alpha } }}{{\frac{d_{{A}{E_j}}^\alpha}{\sum\limits_{k = 1}^N {p_{{A_k}}}}  + \sum\limits_{i = 1}^N {{p_{{A_i}}}^{ - 1}d_{{A_i}{A_{i + 1}}}^\alpha } }}d{x_{{E_j}}}} } \right].
\label{DF_N_approx3}
\end{align}
\hrulefill
\end{figure*}

Then (\ref{DF_N_approx3}) can be turned to
\begin{align}
{\mathcal{P}_{N\_approx2}} = \exp \left[ { - {K_1} {{{\left( {\sum\limits_{k = 1}^N {p_{{A_k}}}}{\sum\limits_{i = 1}^N{\frac{{d_{{A_i}{A_{i + 1}}}^\alpha }}{{{p_{{A_i}}}}}} } \right)}^{\frac{2}{\alpha }}}} } \right],
\label{DF_N_approx4}
\end{align}
where ${K_1} = \pi {\lambda _{{E}}}\Gamma (1 + \frac{2}{\alpha })\Gamma (1 - \frac{2}{\alpha })$ and $\Gamma ( \cdot )$ is the gamma function.

\subsection{Routing Algorithm Based on the SCP Approximations}
In the former subsections, we derive the approximations of the SCP for both colluding and non-colluding eavesdroppers for any given path. In this subsection, we find the path with highest SCP between an arbitrary source and destination.

\subsubsection{Colluding Eavesdroppers Case}
The routing problem depending on (\ref{DF_C_upper}) is still formidable to be solved. We assume that the transmit powers of all nodes are the same. Then (\ref{DF_C_upper}) can be simplified as
\begin{align}
{\mathcal{P}_{C\_approx}}{{ =  \exp}}\left[ {{{ - }}{{{K_2\left( N \right)}}}{{{\left( {\sum\limits_{i = 1}^N {{{d_{{A_i}{A_{i + 1}}}^\alpha }}} } \right)}^{\frac{2}{\alpha }}}}} \right],
\label{DF_C_upper2}
\end{align}
where ${{{K}}_2}\left( N \right){{  =    }}{ \lambda _{{E}}}\frac{{{\pi }\Gamma ({1-\frac{2}{\alpha}} )\Gamma (\frac{2}{\alpha} + N)}}{{\Gamma (N)}}$.

Based on (\ref{DF_C_upper2}), the secure routing problem for finding the highest SCP path can be expressed as
\begin{align}
\mathop {\max}\limits_{L \in {L_A}_{_S{A_D}}} \exp \left[ {{{ - }}{{{K}}_2\left( \left| L \right| \right)}{{\left( {\sum\limits_{i \in L} {d_{{A_i}{A_{i + 1}}}^\alpha} } \right)}^{\frac{2}{\alpha}}}} \right],
\label{RA_upper_1}
\end{align}
where ${L_{{A_S}{A_D}}}$ is the set of all paths $L$ connecting the pair of source node ${A_S}$ and destination node ${A_D}$. Then (\ref{RA_upper_1}) is equivalent to
\begin{align}
\mathop {\min }\limits_{L \in {L_A}_{_S{A_D}}} {{{K}}_2\left( \left| L \right| \right)}{\left( {\sum\limits_{i \in L} {d_{{A_i}{A_{i + 1}}}^\alpha} } \right)^{\frac{2}{\alpha}}}.
\label{RA_upper_2}
\end{align}

It can be easily shown that (\ref{RA_upper_2}) can be solved by exhaustive search, but computationally expensive. The routing metric of problem (\ref{RA_upper_2}) is not isotonic and the problem cannot be solved easily. However, we can prove that the problem (\ref{RA_upper_2}) can be solved exactly optimally in polynomial time. In the following we detail the process.

Since $\left| L \right|$ can only take the value $1,2, \ldots ,{N_L} - 1$, where the ${N_L}$ is the number of the legitimate nodes. According to the divide-and-conquer principle \cite{Wolsey1998}, then problem (\ref{RA_upper_2}) can be rewritten as\cite{Saad2014}
\begin{align}
{{M}_t}({L^*}) = \mathop {\min}\limits_{1 \le v \le N_L - 1} {{M}_t}({L_v}),
\label{RA_upper_3}
\end{align}
where
\begin{eqnarray}
\nonumber{{M}_t}({L_v}) &=& \mathop {\min}\limits_{L \in {L_A}_{_S{A_D}}:\left| L \right| = v} {{{K}}_2\left( \left| L \right| \right)}{\left( {\sum\limits_{i \in L} {d_{{A_i}{A_{i + 1}}}^\alpha } } \right)^{\frac{2}{\alpha }}}\\
\quad \;\,{\kern 1pt}  &=& \mathop {\min}\limits_{L \in {L_A}_{_S{A_D}}:\left| L \right| = v} {{{K}}_2\left( v \right)}{\left( {\sum\limits_{i \in L} {d_{{A_i}{A_{i + 1}}}^\alpha } } \right)^{\frac{2}{\alpha }}}.
\label{RA_upper_4}
\end{eqnarray}
Here ${L^*}$ and ${L_v}$ are the optimal solution to problem (\ref{RA_upper_2}) and subproblem (\ref{RA_upper_4}), respectively; ${{M}_t}({L^*})$ and ${{M}_t}({L_v})$ are the corresponding optimal values of the objective function.

We can solve each subproblem (\ref{RA_upper_4}) to get the optimal solution to problem (\ref{RA_upper_2}). But the subproblem (\ref{RA_upper_4}) is still arduous to be solved, we relax it to
\begin{align}
{{M}_t}({\tilde L_v}) = \mathop {\min}\limits_{L \in {L_A}_{_S{A_D}}:\left| L \right| \le v} {{{K}}_2\left( v\right)}{\left( {\sum\limits_{i \in L} {d_{{A_i}{A_{i + 1}}}^\alpha } } \right)^{\frac{2}{\alpha }}},
\label{RA_upper_5}
\end{align}
where ${\tilde L_v}$ and ${{M}_t}({\tilde L_v})$ denote the optimal solution and the corresponding optimal value of the objective function to the relaxed problem (\ref{RA_upper_5}), respectively. In the following, we discuss the relationship between problem (\ref{RA_upper_2}) and (\ref{RA_upper_5}).

According to (\ref{RA_upper_3}) and (\ref{RA_upper_4}), we can obtain
\begin{align}
{{M}_t}({L^*}) = \mathop {\min}\limits_{1 \le v \le N_L - 1} {{{K}}_2\left( v \right)}{\left( {\sum\limits_{i \in {L_v}} {d_{{A_i}{A_{i + 1}}}^\alpha } } \right)^{\frac{2}{\alpha }}}.
\label{RA_upper_6}
\end{align}
Since (\ref{RA_upper_5}) is the relaxed problem of (\ref{RA_upper_4}) and ${L_v}$ is also a feasible solution to (\ref{RA_upper_5}), then
\begin{eqnarray}
\nonumber{{M}_t}({L^*}) &\ge & \mathop {\min}\limits_{1 \le v \le N_L - 1} {{M}_t}({{\tilde L}_v})\\
\,\,\quad  &=& \mathop {\min}\limits_{1 \le v \le N_L - 1} {{{K}}_2\left( v \right)}{\left( {\sum\limits_{i \in {{\tilde L}_v}} {d_{{A_i}{A_{i + 1}}}^\alpha } } \right)^{\frac{2}{\alpha }}}.
\label{RA_upper_7}
\end{eqnarray}
It can be easily known that $\left| {{{\tilde L}_v}} \right| \le v$ since ${\tilde L_v}$ is the optimal solution to the relaxed problem (\ref{RA_upper_5}) and ${{{K}}_2\left( \left| {{{\tilde L}_v}} \right| \right)} \le {{{K}}_2\left( v \right)}$, then
\begin{align}
{{M}_t}({L^*}) \ge \mathop {\min}\limits_{1 \le v \le N_L - 1} {{{K}}_2\left( \left| {{{\tilde L}_v}} \right| \right)}{\left( {\sum\limits_{i \in {{\tilde L}_v}} {d_{{A_i}{A_{i + 1}}}^\alpha } } \right)^{\frac{2}{\alpha }}}.
\label{RA_upper_8}
\end{align}
Since ${\tilde L_v}$ is also a feasible solution to problem (\ref{RA_upper_4}), then
\begin{align}
{{M}_t}({L^*}) \le \mathop {\min}\limits_{1 \le v \le N_L - 1} {{{K}}_2\left( \left| {{{\tilde L}_v}} \right| \right)}{\left( {\sum\limits_{i \in {{\tilde L}_v}} {d_{{A_i}{A_{i + 1}}}^\alpha } } \right)^{\frac{2}{\alpha }}}.
\label{RA_upper_9}
\end{align}
From (\ref{RA_upper_8}) and (\ref{RA_upper_9}), we can easily obtain
\begin{align}
{{M}_t}({L^*}) = \mathop {\min}\limits_{1 \le v \le N_L - 1} {\tilde {M}_t}({\tilde L_v})
\label{RA_upper_10},
\end{align}
and
\begin{align}
{\tilde {M}_t}({\tilde L_v}) = {{{K}}_2\left( \left| {{{\tilde L}_v}} \right| \right)}{\left( {\sum\limits_{i \in {{\tilde L}_v}} {d_{{A_i}{A_{i + 1}}}^\alpha } } \right)^{\frac{2}{\alpha }}}.
\label{RA_upper_11}
\end{align}
(\ref{RA_upper_10}) and (\ref{RA_upper_11}) imply that problem (\ref{RA_upper_2}) can be solved optimally by solving a sequence of relaxed subproblems (\ref{RA_upper_5}).  Based on the fact that the path loss exponent $\alpha  > 2$,  it is easy to know that the solution to the relaxed subproblem (\ref{RA_upper_5}) for a given hop-count $v$ is equivalent to $\mathop {\min}\limits_{L \in {L_A}_{_S{A_D}}:\left| L \right| \le v} \sum\limits_{i \in L} {d_{{A_i}{A_{i + 1}}}^\alpha }$, which means that each link uses $d_{{A_i}{A_{i + 1}}}^\alpha$ as the link weights to find the path connecting source node ${A_S}$ and destination node ${A_D}$ which has the minimum total link weights and is no more than $v$ hops. The problem can be directly solved by the classical Bellman-Ford shortest path algorithm which computes shortest paths from a single source vertex to all of the other vertices in a weighted digraph. A distributed variant of the algorithm is used in distance-vector routing protocols, for example the Routing Information Protocol (RIP) \cite{bertsekas1992data}. However, the number of hops ${\left| L \right|}$ in the objective function in (\ref{RA_upper_2}) changes with the selected path $L$. Having the weighting factor of ${\left| L \right|}$ in the objective function, the optimization problem cannot be solved directly by using the classical Bellman-Ford algorithm, because it does not take the weighting factor into account. Hence, we develop a revised Bellman-Ford algorithm as shown in Algorithm \ref{alg:al1} below. The classical Bellman-Ford algorithm has an implicit property that at its $h$th iteration, it identifies the optimal path from the source to the destination among all paths of at most $h$ hops. This property is used in Step 1 of the algorithm. On the other hand, Steps 2 and 3 reflect our revision in the Bellman-Ford algorithm in order to solve the problem in (\ref{RA_upper_2}). The whole procedure is shown in Algorithm \ref{alg:al1}.
\begin{algorithm}[h]
\caption{The routing algorithm for the colluding eavesdroppers case.}
\label{alg:al1}
\begin{algorithmic}[1]
\Require
The transmission distance $d_{{A_i}{A_{i + 1}}}$ between the legitimate nodes;
\Ensure
\State Each legitimate nodes use $d_{{A_i}{A_{i + 1}}}^\alpha$ as link weights , obtain the shortest path ${\tilde L_v}$ in each iteration $v\left( {1, \ldots ,N_L-1} \right)$ by the classical Bellman-Ford shortest path algorithm;
\State Calculate the function values for each path ${\tilde L_v}$ using (\ref{RA_upper_11});
\State Get the optimal path ${L^*}$ with the minimum function value using (\ref{RA_upper_10});\\
\Return ${L^*}$;
\end{algorithmic}
\end{algorithm}

Before using the algorithm, each legitimate node calculates the distances between itself and all other nodes in the network and stores the topology information which contains the neighbor list and transmission distance $d_{{A_i}{A_{i + 1}}}$ between them. Then it sends its topology information to all neighboring nodes. Note that the value of $\lambda_E$ does not influence the routing algorithm, since SCP decreases as the value of $\lambda_E$ increases as shown in the exact expression (12). The optimal secure path will always have the highest SCP which is independent with the value of $\lambda_E$. The proposed routing algorithm provides a theoretical basis for finding a link weight $d_{{A_i}{A_{i + 1}}}^\alpha$, which is the key point of a routing algorithm, considering the security. Without the proposed algorithm, the classical Bellman-Ford algorithm does not have a reasonable way to choose a link weight which takes the security into consideration. The complexity of the classical Bellman-Ford algorithm is $O({N_L}^3)$ \cite{bertsekas1992data}. From Algorithm \ref{alg:al1}, it is clear that the computational complexity is dominated by Step 1. Hence, our proposed algorithm has the same level of computational complexity as the classical Bellman-Ford algorithm, which is $O({N_L}^3)$. It is polynomial and much lower than that of the exhaustive search whose complexity is $O(\left({N_L-2}\right)!)$.

\subsubsection{Non-colluding Eavesdroppers Case}
Depending on the SCP approximation (\ref{DF_N_approx4}), the highest SCP path can be presented as the following problem:
\begin{align}
\mathop {\max}\limits_{L \in {L_A}_{_S{A_D}}} \exp \left[ { - {K_1}{{{\left( {{\sum\limits_{k \in L} {p_{{A_k}}}}\sum\limits_{i \in L} {\frac{{d_{{A_i}{A_{i + 1}}}^\alpha }}{{{p_{{A_i}}}}}} } \right)}^{\frac{2}{\alpha }}}} } \right],
\label{N_Approx1}
\end{align}
where ${L_{{A_S}{A_D}}}$ is the set of all paths $L$ connecting the pair of nodes (${A_S}$, ${A_D}$). When the network parameters ${\lambda _{{E}}}$ and $\alpha$ are determined, ${K_1}$ is a constant and positive. Then (\ref{N_Approx1}) is equivalent to
\begin{align}
\mathop {\min}\limits_{L \in {L_A}_{_S{A_D}}}  {{{\left( {{\sum\limits_{k \in L} {p_{{A_k}}}}\sum\limits_{i \in L} {\frac{{d_{{A_i}{A_{i + 1}}}^\alpha }}{{{p_{{A_i}}}}}} } \right)}^{\frac{2}{\alpha }}}}.
\label{N_Approx2}
\end{align}

We assume that the transmit powers of all nodes are the same. Then (\ref{N_Approx2}) can be simplified as
\begin{align}
\mathop {\min}\limits_{L \in {L_A}_{_S{A_D}}}{\left(\left| L \right| {\sum\limits_{i \in L} {d_{{A_i}{A_{i + 1}}}^\alpha } } \right)^{\frac{2}{\alpha }}}.
\label{N_Approx3}
\end{align}

Similar to the case of colluding eavesdroppers, we also can prove that problem (\ref{N_Approx3}) can be solved optimality by solving a sequence of subproblems
\begin{align}
{{M}_u}({\mathord{\buildrel{\lower3pt\hbox{$\scriptscriptstyle\frown$}}
\over L} _u}) = \mathop {\min}\limits_{L \in {L_A}_{_S{A_D}}:\left| L \right| \le u}{\left(u {\sum\limits_{i \in L} {d_{{A_i}{A_{i + 1}}}^\alpha} } \right)^{\frac{2}{\alpha}}},
\label{N_Approx4}
\end{align}
\begin{align}
{{M}_u}(\mathord{\buildrel{\lower3pt\hbox{$\scriptscriptstyle\frown$}}
\over L} _u^*) = \mathop {\min}\limits_{1 \le u \le N_L - 1} {\tilde {M}_u}({\mathord{\buildrel{\lower3pt\hbox{$\scriptscriptstyle\frown$}}
\over L} _u}),
\label{N_Approx5}
\end{align}
and
\begin{align}
{\tilde {M}_u}({\mathord{\buildrel{\lower3pt\hbox{$\scriptscriptstyle\frown$}}
\over L} _u}) ={\left({\left| {{{\mathord{\buildrel{\lower3pt\hbox{$\scriptscriptstyle\frown$}}
\over L} }_u}} \right|} {\sum\limits_{i \in {{\mathord{\buildrel{\lower3pt\hbox{$\scriptscriptstyle\frown$}}
\over L} }_u}} {d_{{A_i}{A_{i + 1}}}^\alpha} } \right)^{\frac{2}{\alpha}}},
\label{N_Approx6}
\end{align}
where $\mathord{\buildrel{\lower3pt\hbox{$\scriptscriptstyle\frown$}}\over L} _u^*$ and ${\mathord{\buildrel{\lower3pt\hbox{$\scriptscriptstyle\frown$}}
\over L} _u}$ are the optimal solution to problem (\ref{N_Approx3}) and subproblem (\ref{N_Approx4}), respectively; ${{M}_u}(\mathord{\buildrel{\lower3pt\hbox{$\scriptscriptstyle\frown$}}\over L} _u^*)$ and ${{M}_u}({\mathord{\buildrel{\lower3pt\hbox{$\scriptscriptstyle\frown$}}\over L} _u})$ are the corresponding optimal values of the objective function; ${\tilde {M}_u}({\mathord{\buildrel{\lower3pt\hbox{$\scriptscriptstyle\frown$}}\over L} _u})$ is the function value of (\ref{N_Approx6}). The whole procedure is shown in Algorithm \ref{alg:al2}.

\begin{algorithm}[h]
\caption{The routing algorithm for the non-colluding eavesdroppers case.}
\label{alg:al2}
\begin{algorithmic}[1]
\Require
The transmission distance $d_{{A_i}{A_{i + 1}}}$ between the legitimate nodes;
\Ensure
\State Each legitimate nodes use $d_{{A_i}{A_{i + 1}}}^\alpha$ as link weights , obtain the shortest path ${\mathord{\buildrel{\lower3pt\hbox{$\scriptscriptstyle\frown$}}
\over L} _u}$ in each iteration $u\left( {1, \ldots ,N_L-1} \right)$ by the classical Bellman-Ford shortest path algorithm;
\State Calculate the function values for each path ${\mathord{\buildrel{\lower3pt\hbox{$\scriptscriptstyle\frown$}}
\over L} _u}$ using (\ref{N_Approx6});
\State Get the optimal path $\mathord{\buildrel{\lower3pt\hbox{$\scriptscriptstyle\frown$}}\over L} _u^*$ with the minimum function value using (\ref{N_Approx5});\\
\Return $\mathord{\buildrel{\lower3pt\hbox{$\scriptscriptstyle\frown$}}\over L} _u^*$;
\end{algorithmic}
\end{algorithm}

The computational complexity of Algorithm \ref{alg:al2} for the non-colluding eavesdroppers case is the same as Algorithm \ref{alg:al1} for the colluding eavesdroppers case which is also $O({N_L}^3)$.

\section{Numerical Results and Discussion}
In this section, we present numerical results and evaluate the performance of the derived expressions of SCP, then we compare the performance of different routing algorithms on security. We take path loss exponent $\alpha  = 4$, and we assume that all the transmit powers are the same.

\subsection{Performance of Derived SCP}
We simulate a multihop wireless network, in which the nodes are deployed in a $2000 \times 2000$ square area. The eavesdroppers are located at random positions which follow a homogeneous PPP. In this subsection, we consider an example of 6 legitimate nodes ${A_1} \sim {A_6}$, and they locate at $\left(- 10,0\right)$, $\left(5\cos\left(0.75\pi\right),5\sin\left(0.75\pi\right)\right)$, $\left(0,0 \right)$, $\left(5\cos\left(- 0.25\pi\right),5\sin\left(- 0.25\pi\right)\right)$, $\left(10,0\right)$ and $\left(15\cos\left(0.25\pi\right),15\sin\left(0.25\pi\right) \right)$. It takes 10000 simulation runs to obtain Monte Carlo simulation results.

\begin{figure}[t]
\centering
  \includegraphics[width=8cm]{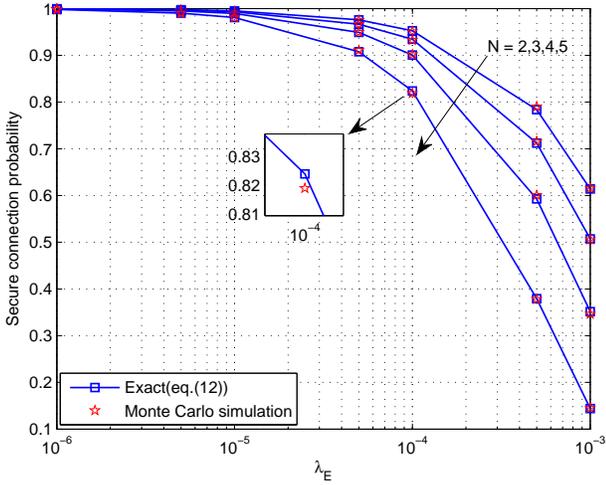}\\
  \caption{Monte Carlo simulation results of SCP for colluding eavesdroppers case. The squares represent (\ref{DF_C_3}). The stars show the Monte Carlo simulation of (\ref{DF_C_1}). }\label{fig:M_DF_C}
\end{figure}

Fig. \ref{fig:M_DF_C} depicts the Monte Carlo simulation results of SCP for different ${\lambda _E}$. It can be seen that our analysis results match with the Monte Carlo simulation results, which validates our analysis.
\begin{figure}[t]
\centering
  \includegraphics[width=8cm]{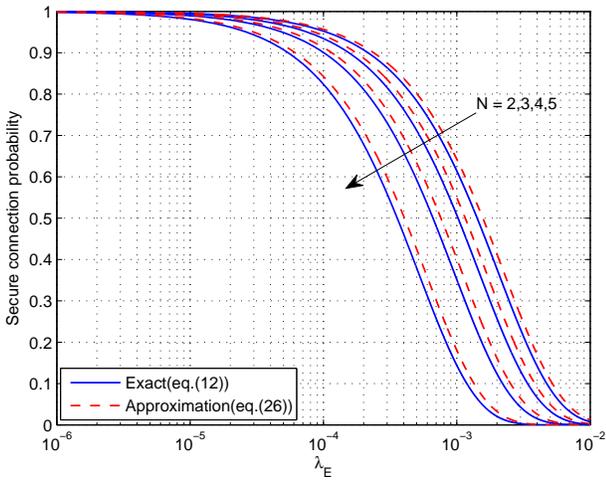}\\
  \caption{SCP for colluding eavesdroppers case. The solid lines represent (\ref{DF_C_3}) and the dashed lines denote the SCP approximation (\ref{DF_C_upper2}). }\label{fig:DF_C_upper}
\end{figure}

Fig. \ref{fig:DF_C_upper} illustrates the SCP for the case of colluding eavesdroppers as a function of ${\lambda _E}$. As the value of ${\lambda _E}$ and the number of hops grow, the SCP decreases. The gap between the approximation and the exact value of the SCP is small, and we can see that our SCP approximation is a precise approximation of the exact value for all ${\lambda _E}$.

\begin{figure}[t]
\centering
  \includegraphics[width=8cm]{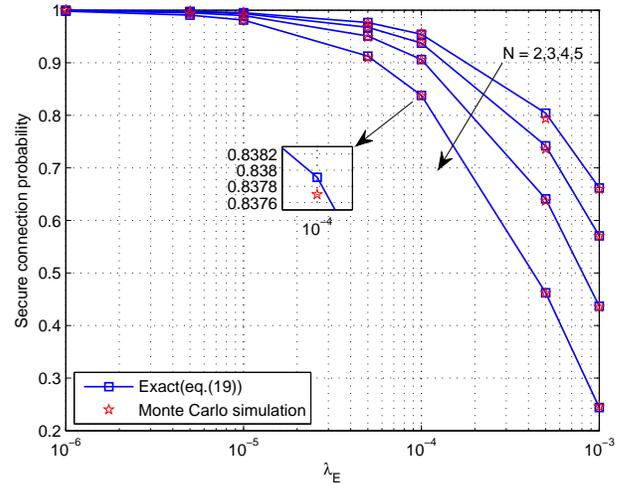}\\
  \caption{Monte Carlo simulation results of SCP for non-colluding eavesdroppers case. The squares represent (\ref{DF_N_4}). The stars show the Monte Carlo simulation of (\ref{DF_N_1}).}\label{fig:M_DF_N}
\end{figure}

Fig. \ref{fig:M_DF_N} depicts the Monte Carlo simulation results of SCP for the case of non-colluding eavesdroppers as a function of ${\lambda _E}$. Again, we see that the analytical results match well with the simulation.

\begin{figure}[t]
\centering
  \includegraphics[width=8cm]{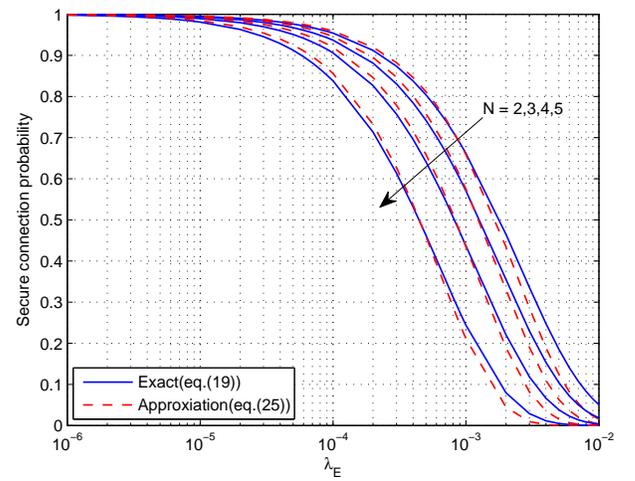}\\
  \caption{SCP for non-colluding eavesdroppers case. The solid lines represent (\ref{DF_N_4}) and the dashed lines denote SCP approximation (\ref{DF_N_approx4}). }\label{fig:DF_N_approx}
\end{figure}
Fig. \ref{fig:DF_N_approx} illustrates the SCP for the case of non-colluding eavesdroppers as a function of ${\lambda _E}$. We can see that the SCP approximation (\ref{DF_N_approx4}) is accurate compared to the exact value obtained in (\ref{DF_N_4}) for a wide range of ${\lambda _E}$. This implies that the accuracy of the approximation is good for a wide range of eavesdropper density. Hence, the derived routing algorithm based on the approximation will give the optimal result in most cases.

\subsection{Performance of Routing Algorithm}
We consider a multihop wireless network in which $N_L = 32$ legitimate nodes are placed uniformly at random on a $50 \times 50$ square area in the center of the network. The source node is placed at the lower left corner of the network and the destination is located at the upper right corner. Note that the eavesdroppers are still randomly distributed in the entire network of size $2000 \times 2000$. Our goal is to find the route that gives the highest SCP between the source and destination. For comparison, we consider the optimal route from exhaustive search as the benchmark routing algorithm.
\begin{figure}[t]
\centering
  \includegraphics[width=8cm]{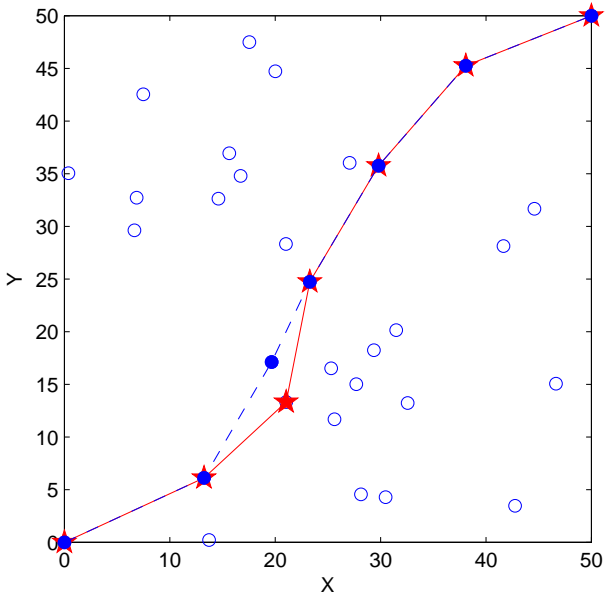}\\
  \caption{Routing algorithm based on the SCP approximation (\ref{DF_C_upper2}) under the case of colluding eavesdroppers. A snapshot of the network is shown when $N_L = 32$ legitimate nodes (shown by circles) are placed uniformly at random. The proposed route is plotted by the red solid line and the benchmark route is shown by the blue dashed line.}\label{route1}
\end{figure}

In Fig. \ref{route1}, we present a snapshot of the network for the case of colluding eavesdroppers. The proposed route based on the SCP approximation (\ref{DF_C_upper2}) and the benchmark route by exhaustive search between the source and destination are plotted in the picture. The link weight is $d_{{A_i}{A_{i + 1}}}^\alpha$. The actual source-destination SCP of the proposed route and benchmark route computed by (\ref{DF_C_3}) for different $\lambda_E$ are shown in Table \ref{tab:route}. It can be seen that our proposed route is exceedingly close to the benchmark route on security.
\begin{table}[h]
\centering
\caption{SCP of the Routing Algorithm under the Case of Colluding Eavesdroppers for Different $\lambda_E$.}
\label{tab:route}
\begin{tabular}{|c|c|c|c|}
\hline
$\lambda_E$     &$10^{-6}$  &$10^{-5}$  &$10^{-4}$  \\ \hline
proposed route  &0.9933 &0.9349 &$0.5103$  \\ \hline
benchmark route &0.9933 &0.9351 &$0.5112$  \\ \hline
\end{tabular}
\end{table}

\begin{figure}[t]
\centering
  \includegraphics[width=8cm]{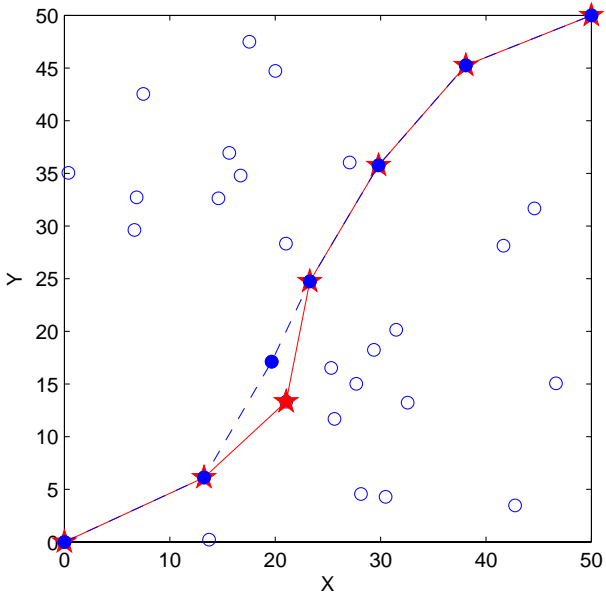}\\
  \caption{Routing algorithm based on the SCP approximation (\ref{DF_N_approx4}) under the case of non-colluding eavesdroppers. A snapshot of the network with the same system nodes as in Fig. \ref{route1} is shown. The proposed route is plotted by the red solid line and the benchmark route is shown by the blue dashed line.}\label{route2_f}
\end{figure}

In Fig. \ref{route2_f}, we present a snapshot of the route based on the SCP approximation (\ref{DF_N_approx4}) with the same system nodes as in Fig. \ref{route1} under the case of the non-colluding eavesdroppers. As shown in the figure, we can derive the same results as the case of colluding eavesdroppers. Specially, the optimal route for the non-colluding case is the same as that for the colluding eavesdroppers. This is because that the eavesdropper with the strongest signal reception contributes the most in the eavesdropping capability of a set of colluding eavesdroppers, unless the density of eavesdroppers becomes comparable to that of the legitimate nodes. This implies that in most scenarios the best secure route against the strongest eavesdropper, which is in fact the non-colluding case, is also likely to be the best route against all eavesdroppers when they collude. The actual source-destination SCP of the proposed route and benchmark route computed by (\ref{DF_N_4}) for different $\lambda_E$ are shown in Table \ref{tab:route2}.
\begin{table}[h]
\centering
\caption{SCP of the Routing Algorithm under the Case of Non-colluding Eavesdroppers for Different $\lambda_E$.}
\label{tab:route2}
\begin{tabular}{|c|c|c|c|}
\hline
$\lambda_E$     &$10^{-6}$  &$10^{-5}$  &$10^{-4}$  \\ \hline
proposed route  &0.9933 &0.9373 &$0.5651$  \\ \hline
benchmark route &0.9934 &0.9375 &$0.5662$  \\ \hline
\end{tabular}
\end{table}

We assume ${\lambda _E} = {10^{ - 5}}$. For comparison, we consider the optimal route from exhaustive search as the benchmark routing algorithm. For different number of legitimate nodes, we simulate the routing algorithms $1000$ times based on the SCP approximation (\ref{DF_C_upper2}) and exhaustive search. However, the computational complexity of the exhaustive search for the case of non-colluding eavesdroppers is too high to simulate, we only show the case of colluding eavesdroppers in the following. Note that enumerating all the routes of the benchmark routing algorithm from the source to the destination becomes prohibitive in a large number of legitimate nodes, so we only simulate the number of the legitimate nodes up to 12. The results are shown in Table \ref{route2}.
\begin{table*}[!t]
\caption {Comparison of Different Routing Algorithms Varying with the Number of Legitimate Nodes }
\label{route2}
\centering
\begin{tabular}{|l|c|c|c|c|c|c|c|c|c|}
\hline
$N_L$       &4       &5      &6       &7      &8      &9      &10      &11      &12     \\ \hline
$\mathcal{P}_{SC\_best}$     &0.8364 &0.8522 &0.8635 &0.8731 &0.8794 &0.8847 &0.8910 &0.8949 &0.8986 \\ \hline
$\mathcal{P}_{SC\_approx}$    &0.8360 &0.8518 &0.8632 &0.8728 &0.8790 &0.8844 &0.8908 &0.8946 &0.8983 \\ \hline
$\mathcal{P}_{EQ\_approx}$    &91.4\% &90.9\% &88.1\% &87.3\% &85.6\% &84.3\% &85.0\% &83.3\% &80.8\% \\ \hline
\end{tabular}
\end{table*}

In Table \ref{route2}, $N_L$ denotes the number of the legitimate nodes. $\mathcal{P}_{SC\_approx}$ and $\mathcal{P}_{SC\_best}$ represent the exact SCP of the route for the approximation and exhaustive search, respectively. $\mathcal{P}_{EQ\_approx}$ represents the probability of the routes based on the SCP approximation which coincide with the benchmark routes. As shown in the table, the SCP increases with the number of legitimate nodes growing. It is because that more legitimate nodes will give more chance to get a safer route for a given source-destination pair of nodes. The gap between the proposed route and benchmark route is minuscule. The probability of the route based on the SCP approximation choosing the same route as benchmark route is $80.8\%\sim91.4\%$. Such a small but notable difference in the routes results in very insignificant performance degradation. As we can see, the route based on the SCP approximation is intensely close to the benchmark route on security.

\section{Conclusion}
This paper studied the secure routing problem in multihop wireless networks. Given a path of a source-destination pair of nodes, we obtained the exact expressions of the secure connection probability (SCP) for both colluding and non-colluding eavesdroppers. Then the SCP approximations were derived to facilitate finding the routing algorithm. Based on the SCP approximations, we solved the routing problem between an arbitrary pair of nodes to find the highest SCP path connecting them. Our proposed secure routing protocol finds the optimal path in a distributed way by using a revised Bellman-Ford algorithm.

Our work focused on a benchmarking scenario where the most commonly-used DF relaying protocol is assumed. To further improve the secrecy performance, the RaF relaying protocol can be implemented which uses independent codewords at the relays and is specifically designed from the viewpoint of physical layer security. In our future work, we will extend our analysis to this scenario and compare with the benchmarking case to see to what extent the secure routing protocols differ from each other.


%

\appendices
\section{Proof of Lemma \ref{lemma1}}\label{appendices_A}
Let
\begin{align}
{{ }}{{{f}}_n}\left( x \right) = \int_{- \infty }^\infty  {\left( {{{1 - }}\prod\limits_{{{k}} = {{1}}}^n {\frac{{{1}}}{{1 + {B_k}{{(x + {a_k})}^{-2}}}}}} \right)} dx,
\end{align}
\begin{align}
{g_n}\left( x \right) = \int_{ - \infty }^\infty  {\left( {{{1 - }}\prod\limits_{{{k}} = {{1}}}^n {\frac{{{1}}}{{1 + {B_k}{{(x + {a})}^{ - 2}}}}} } \right)} dx.
\end{align}

Let $x=x+a$, then
\begin{align}
{g_n}\left( x \right) = \int_{ - \infty }^\infty  {\left( {{{1 - }}\prod\limits_{{{k}} = {{1}}}^n {\frac{{{1}}}{{1 + {B_k}{{ x }^{ - 2}}}}} } \right)} dx.
\end{align}

When $n = 1$,
\begin{align}
{{ }}{{{f}}_1}\left( x \right) = {{{g}}_1}\left( x \right) =\sqrt{B_1}\pi.
\end{align}

When $n = 2$,
\begin{align}
{{ }}{{{f}}_2}\left( x \right) = \frac{{\left( {\sqrt B_1  + \sqrt B_2 } \right)\left( {B_1 + \sqrt {{B_1}{B_2}}  + B_2 + {{\mathord{\buildrel{\lower3pt\hbox{$\scriptscriptstyle\frown$}} \over a}}^2}} \right)}}{{B_1 + 2\sqrt {{B_1}{B_2}}  + B_2 + {{\mathord{\buildrel{\lower3pt\hbox{$\scriptscriptstyle\frown$}} \over a}}^2}}},
\end{align}

\begin{align}
{{{g}}_2}\left( x \right) = \frac{{B_1 + \sqrt {{B_1}{B_2}}  + B_2}}{{\sqrt B_1  + \sqrt B_2 }},
\end{align}
where ${\mathord{\buildrel{\lower3pt\hbox{$\scriptscriptstyle\frown$}} \over a}} = a_2- a_1 $, then

\begin{align}
{{{f}}_2}\left( x \right) - &{{{g}}_2}\left( x \right)=&\\
&\nonumber\frac{{{{\mathord{\buildrel{\lower3pt\hbox{$\scriptscriptstyle\frown$}} \over a}}^2}\left( {B_1 + \sqrt {{B_1}{B_2}}  + B_2} \right)}}{{\left( {\sqrt B_1  + \sqrt B_2 } \right)\left( {{{\mathord{\buildrel{\lower3pt\hbox{$\scriptscriptstyle\frown$}} \over a}}^2} + {{\left( {\sqrt B_1  + \sqrt B_2 } \right)}^2}} \right)}} > 0.&
\end{align}

When $n = 3$,
\begin{align}
{{{f}}_3}\left( x \right) = \int_{ - \infty }^\infty  {\left( {{{1 - }}\prod\limits_{{{k}} = {{1}}}^3 {\frac{{{1}}}{{1 + {B_k}{{(x + {a_k})}^{ - 2}}}}} } \right)} dx.
\label{n_3}
\end{align}

Let $x=x+a_3$, then (\ref{n_3}) can be turned to
\begin{align}
{{{f}}_3}\left( x \right) = \int_{ - \infty }^\infty  {\left( {{{1 - }} {\frac{{{1}}}{{1 + {B_3}{x^{ - 2}}}}} \prod\limits_{{{k}} = {{1}}}^2 {\frac{{{1}}}{{1 + {B_k}{{(x + {b_k})}^{ - 2}}}}} } \right)} dx,
\end{align}
where ${b_k = a_k- a_3 (k<3)}$.
\begin{align}
&\nonumber{{{f}}_3}\left( x \right) - {{{g}}_3}\left( x \right) = \int_{ - \infty }^\infty  {\frac{{{1}}}{{1 + {B_3}{x^{ - 2}}}}} \\
&\times{\left( {\prod\limits_{{{k}} = {{1}}}^2{\frac{{{1}}}{{1 + {B_k}{x^{ - 2}}}}} - \prod\limits_{{{k}} = {{1}}}^2{\frac{{{1}}}{{1 + {B_k}{{(x + {b_k})}^{ - 2}}}}}} \right)}  dx.&
\label{n_4}
\end{align}

According to first mean value theorem\cite{Jeffrey2007}, there exists a constant $-\infty\leq\varepsilon_1\leq\infty$ holding the equation
\begin{align}
\nonumber&{{{f}}_3}\left( x \right) - {{{g}}_3}\left( x \right) = {\frac{{{1}}}{{1 + {B_3}{\varepsilon_1 ^{ - 2}}}}} \\
& \times\int_{ - \infty }^\infty  {\left( {\prod\limits_{{{k}} = {{1}}}^2{\frac{{{1}}}{{1 + {B_k}{x^{ - 2}}}}} - \prod\limits_{{{k}} = {{1}}}^2{\frac{{{1}}}{{1 + {B_k}{{(x + {b_k})}^{ - 2}}}}}} \right)} dx.
\label{n_5}
\end{align}
Then (\ref{n_5}) can be rewritten as
\begin{align}
{{ }}{{{f}}_3}\left( x \right) - {{{g}}_3}\left( x \right) = {\frac{{{1}}}{{1 + {B_3}{\varepsilon_1 ^{ - 2}}}}} \left( {{{{f}}_2}\left( x \right) - {{{g}}_2}\left( x \right)} \right) > 0.
\end{align}

We assume that when $n = j$ and
\begin{align}
{{{f}}_j}\left( x \right) - {{{g}}_j}\left( x \right) > 0.
\end{align}

Then when $n = j+1$, we have
\begin{align}
{{{f}}_{j+1}}\left( x \right) = \int_{ - \infty }^\infty  {\left( {{{1 - }}\prod\limits_{{{k}} = {{1}}}^{j+1} {\frac{{{1}}}{{1 + {B_k}{{(x + {a_k})}^{ - 2}}}}} } \right)}dx.
\label{n_6}
\end{align}

Let $x=x+a_{j+1}$, then (\ref{n_6}) can be turned to
\begin{align}
&\nonumber{{{f}}_{j+1}}\left( x \right) =& \\
&\int_{ - \infty }^\infty  {\left( {{{1 - }}{\frac{{{1}}}{{1 + {B_{j+1}}{x^{ - 2}}}}}\prod\limits_{{{k}} = {{1}}}^{j} {\frac{{{1}}}{{1 + {B_k}{{(x + {c_k})}^{ - 2}}}}} } \right)} dx,&
\end{align}
where ${c_k = a_k- a_{j+1} (k<j+1)}$.

\begin{align}
\nonumber&{{{f}}_{j + 1}}\left( x \right) - {{{g}}_{j + 1}}\left( x \right) = \int_{ - \infty }^\infty  {\frac{{{1}}}{{1 + {B_{j+1}}{x^{ - 2}}}}} \\
& \times{\left( {\prod\limits_{{{k}} = {{1}}}^j{\frac{{{1}}}{{1 + {B_{k}}{x^{ - 2}}}}} - \prod\limits_{{{k}} = {{1}}}^j{\frac{{{1}}}{{1 + {B_{k}}{{(x + {c_k})}^{ - 2}}}}}} \right)} dx.
\label{n_7}
\end{align}

Similar to $n = 3$, (\ref{n_7}) can be rewritten to
\begin{align}
{{{f}}_{j + 1}}\left( x \right) - {{{g}}_{j + 1}}\left( x \right) = {\frac{{{1}}}{{1 + {B_{j+1}}{\varepsilon_2 ^{ - 2}}}}} \left( {{{{f}}_j}\left( x \right) - {{{g}}_j}\left( x \right)} \right) > 0.
\end{align}

So we can conclude that $f_n\left( x \right)$ is greater than $g_n\left( x \right)$ for an arbitrary positive integrate random variable $n > 1$.


\ifCLASSOPTIONcaptionsoff
  \newpage
\fi



\bibliographystyle{IEEEtran}
\bibliography{TCOM-arxiv}
\end{document}